\documentclass[prl,preprint,showpacs,amssymb,floatfix,superscriptaddress]{revtex4}
\usepackage{graphicx} \usepackage{subfigure}

\begin{document}

\title{Universal Anisotropy in Force Networks under Shear}

\author{Srdjan Ostojic}  \altaffiliation[Present address: ]{Laboratoire de Physique Statistique, Ecole Normale Sup\'erieure, 24, rue Lhomond, 75231 Paris Cedex 05, France}\affiliation{Institute for
Theoretical Physics, Universiteit van Amsterdam, Valckenierstraat 65,
1018 XE Amsterdam, The Netherlands}

\author{Thijs J. H. Vlugt} \affiliation{Condensed Matter and
  Interfaces (CMI), Utrecht University, P.O.Box 80.000, 3508 TA Utrecht, The Netherlands}

\author{Bernard Nienhuis} \affiliation{Institute for
Theoretical Physics, Universiteit van Amsterdam, Valckenierstraat 65,
1018 XE Amsterdam, The Netherlands}

\date{\today}

\begin{abstract} 
Scaling  properties of  patterns formed  by large  contact  forces are
studied as a function of  the applied shear stress, in two-dimensional
static  packings  generated  from  the  force  network  ensemble.   An
anisotropic finite-size-scaling analysis  shows that the applied shear
does not  affect the universal  scaling properties of  these patterns,
but simply induces different length scales in the principal directions
of the  macroscopic stress tensor.   The ratio of these  length scales
quantifies the anisotropy  of the force networks, and  is found not to
depend on the  details of the underlying contact  network, in contrast
with other properties such as the yield stress.
\end{abstract}

\pacs{45.70.-n, 45.70.Cc, 46.65.+g}
\maketitle

\begin{figure}
\begin{center} 
\begin{minipage}{0.32\linewidth}
\includegraphics [width=0.95\linewidth]{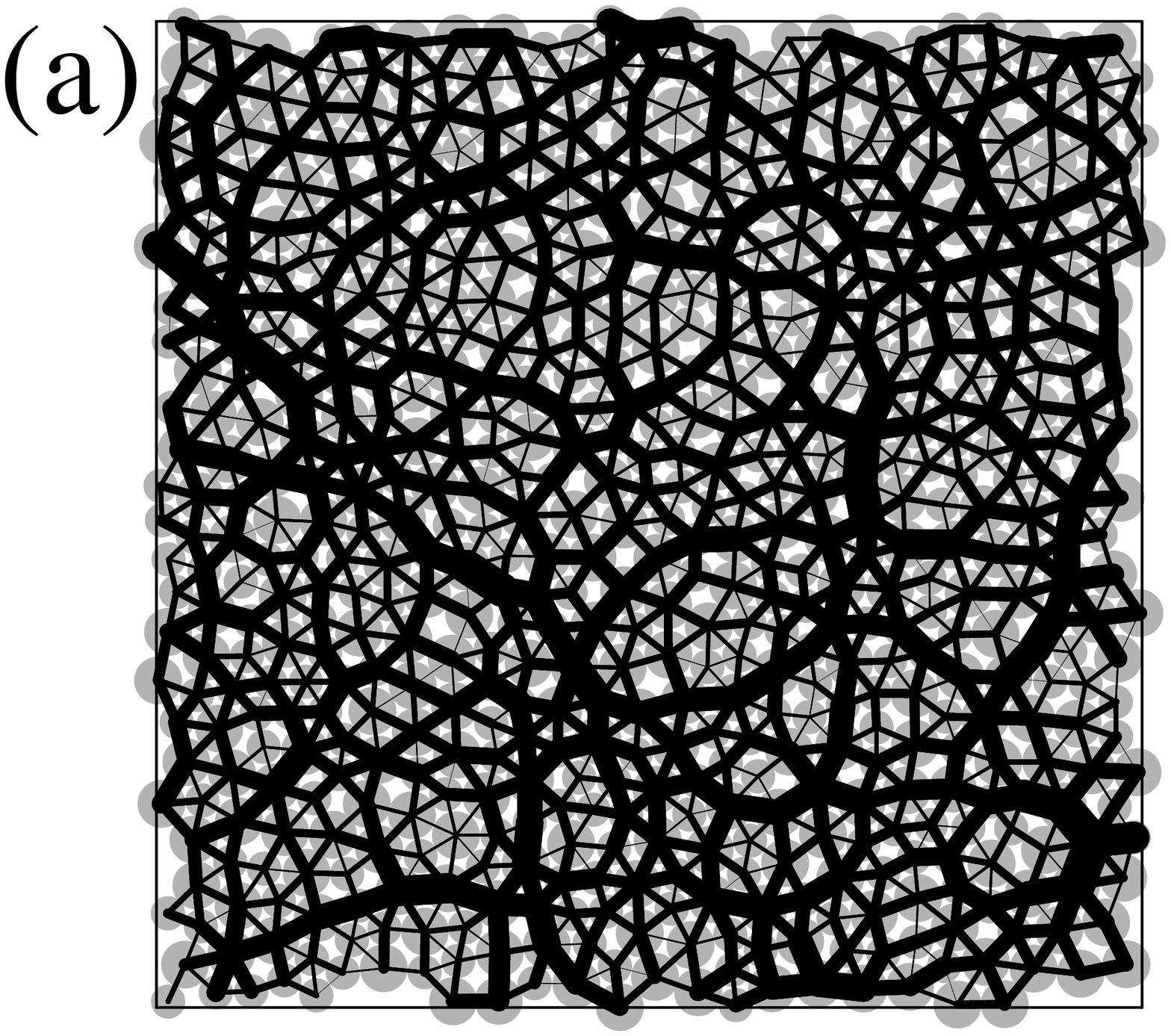}
\end{minipage}
\begin{minipage}{0.32\linewidth}
\includegraphics [width=0.95\linewidth]{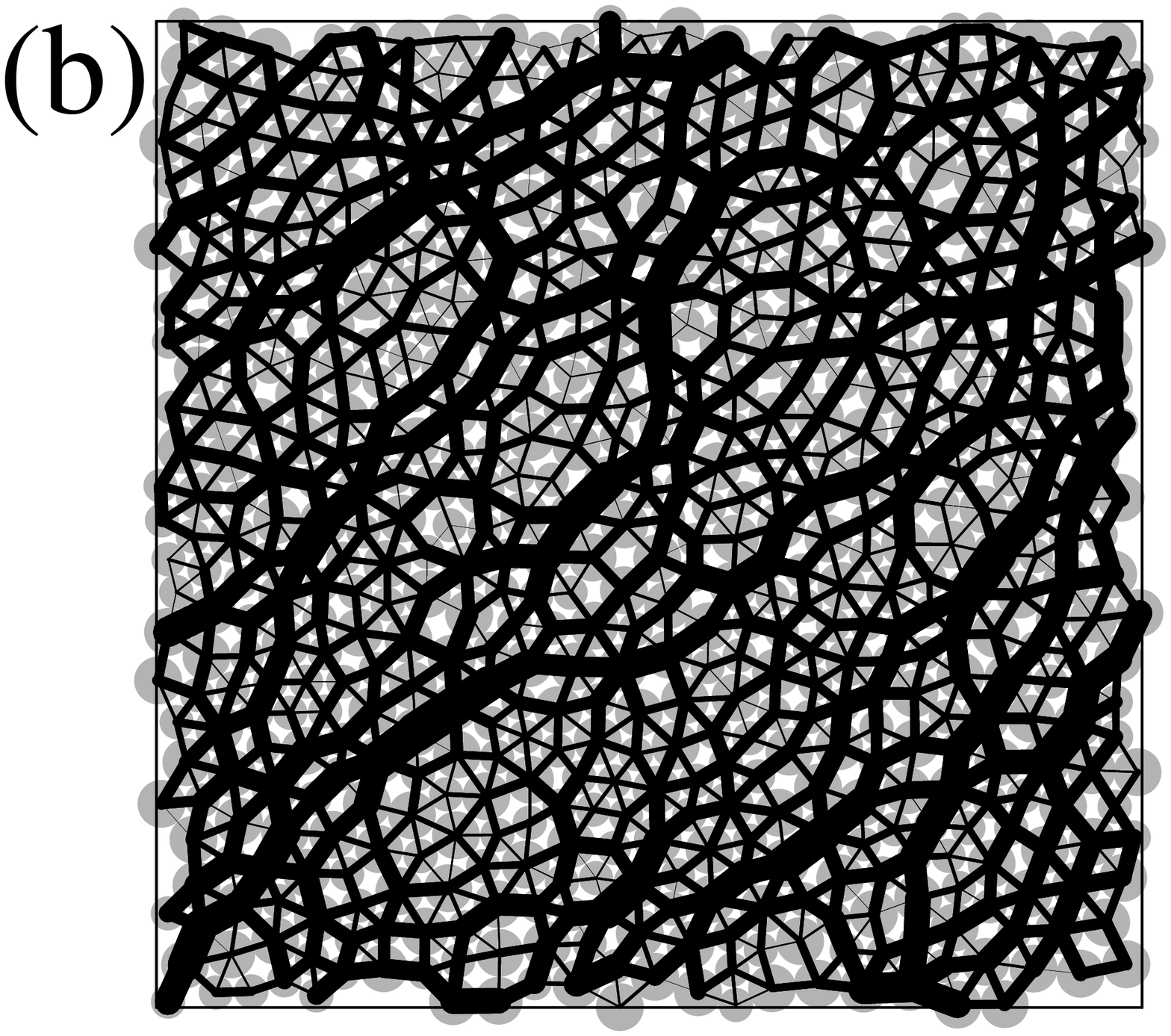}
\end{minipage}
\begin{minipage}{0.32\linewidth}
\includegraphics [width=0.95\linewidth]{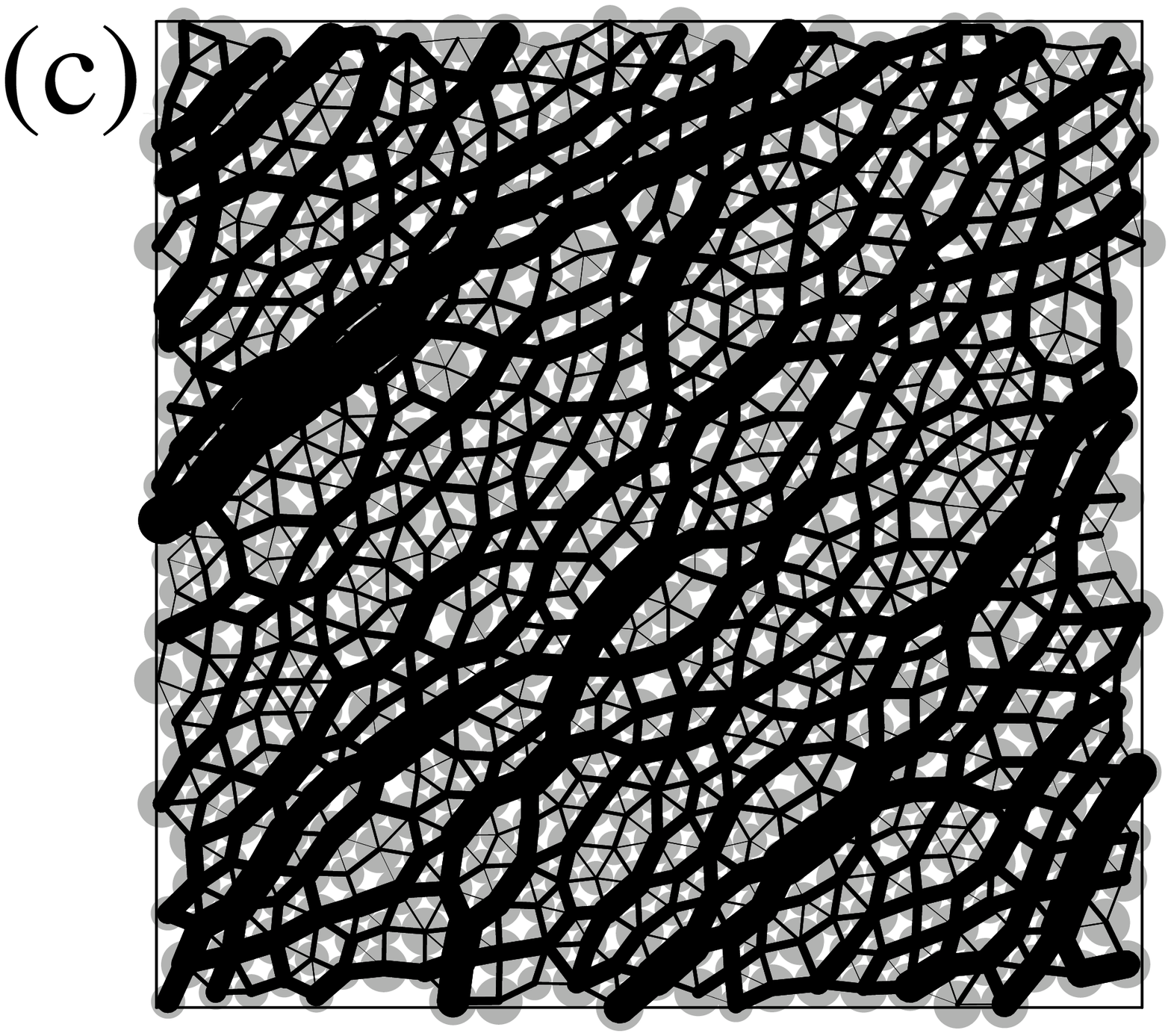}
\end{minipage}

\begin{minipage}{0.32\linewidth}
\includegraphics [width=0.95\linewidth]{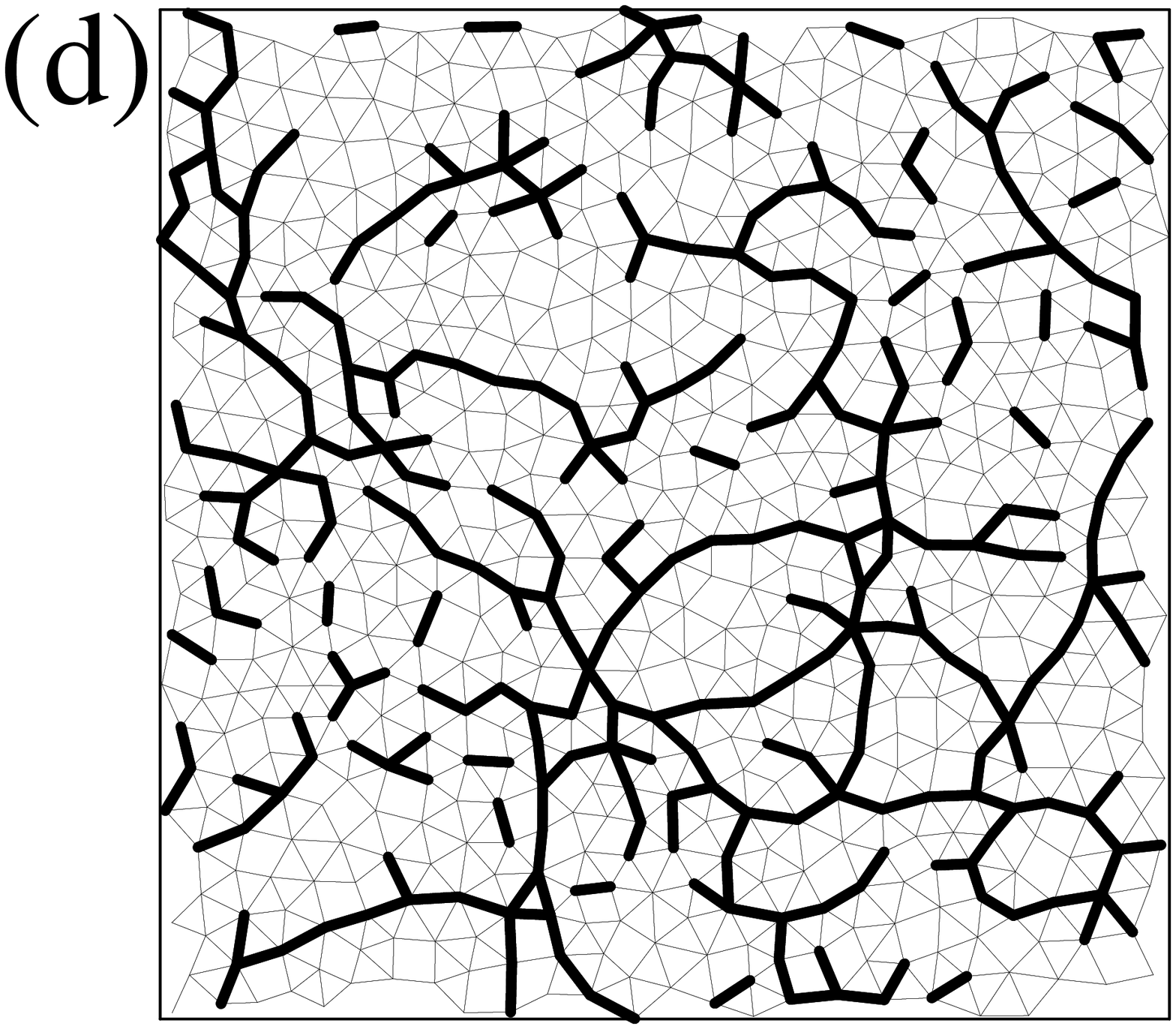}
\end{minipage}
\begin{minipage}{0.32\linewidth}
\includegraphics [width=0.98\linewidth]{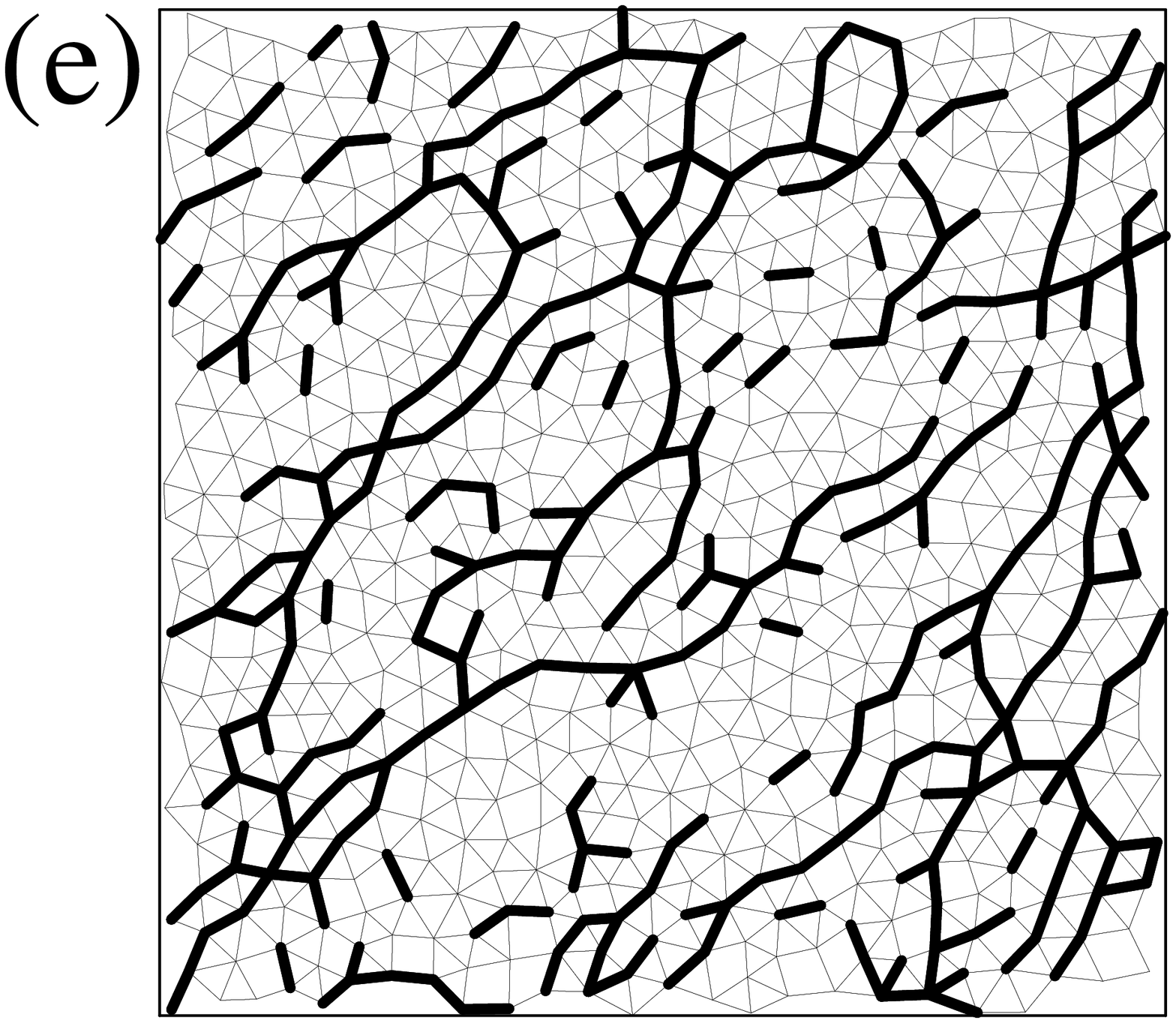}
\end{minipage}
\begin{minipage}{0.32\linewidth}
\includegraphics [width=0.98\linewidth]{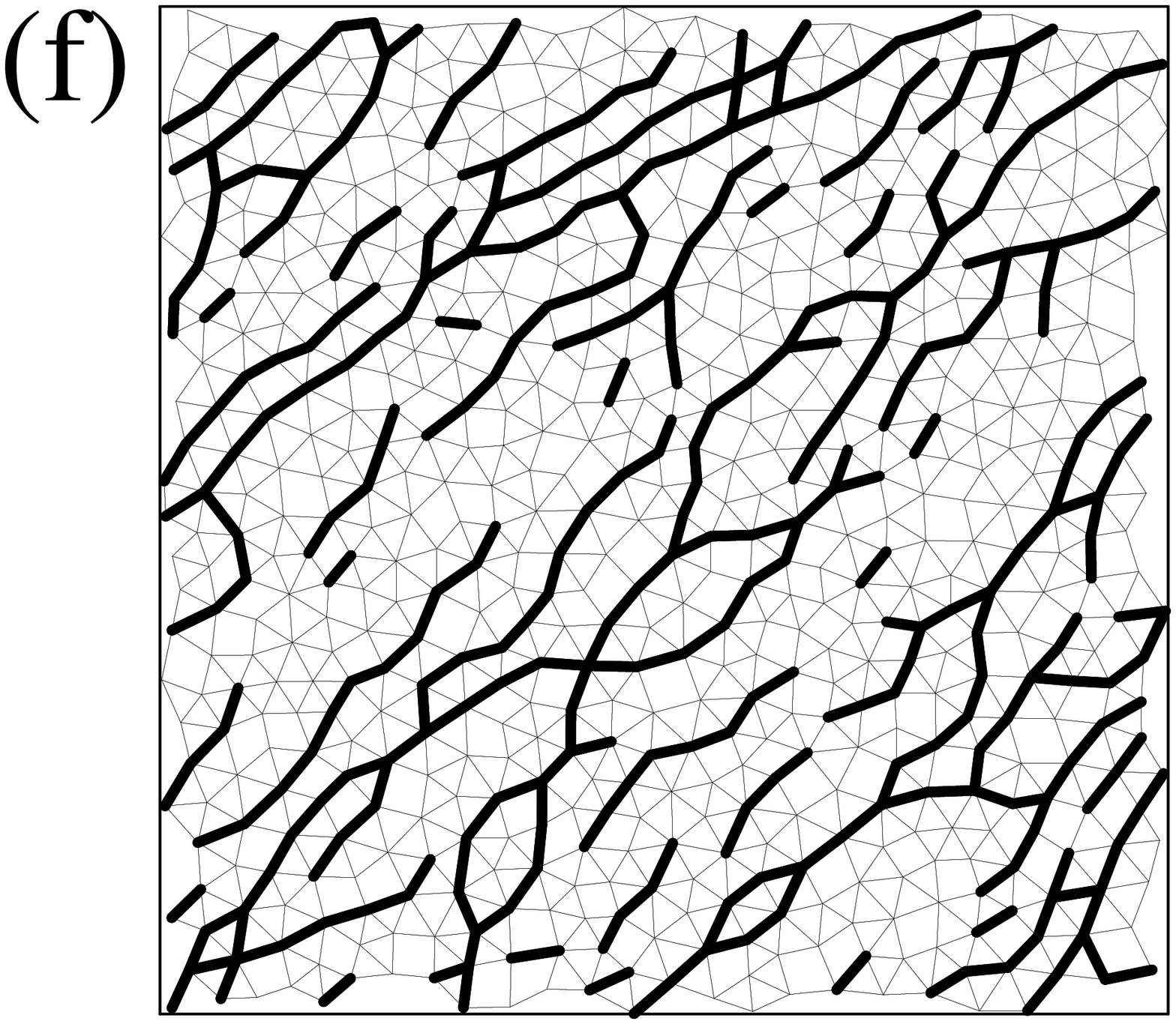}
\end{minipage}
\caption{Effect  of  shear on  force  networks:  (a-c) Force  networks
obtained from the force ensemble, for three different values of shear:
(a)  $\tau=0$,  (b)  $\tau=0.2$,   (c)  $\tau=0.4$.   The  forces  are
represented  as bonds  connecting centers  of grains  of  contact, the
width  of  each  bond  being  proportional to  the  magnitude  of  the
force. The  underlying contact network is  the same for  all values of
$\tau$.  (d-f)  Clusters of forces  larger than a threshold  $f$ (bold
bonds), obtained for $f$ close to the critical value.  \label{fig1}}
\end{center}
\end{figure}

Aggregates of macroscopic particles  such as granular materials, foams
and emulsions are often found  in a disordered, solid-like state whose
mechanical properties  have attracted  much attention in  recent years
\cite{jaeger:rev,degennes:rev,bouch:rev,liu98,jamming:book,radjai98,geng03,majmudar05,atman05-2,snoeijer04-1,snoeijer05,tighe05,ostojic06}.
Under  external  stresses,  these  systems present  a  non-zero  yield
threshold  which is  solely due  to  the intricate  network formed  by
contact forces between particles. One of the most remarked features of
these highly disordered {\em force networks} is the tendency for large
forces to align and form  branching {\em force chains}. When these appear  in  response to an external  stress, such as  a global shear, their spatial inhomogeneity is most striking \cite{radjai98,geng03,majmudar05}.
As the applied shear stress is increased, the large forces orient in a
preferred  direction, and  the  force network  becomes  more and  more
anisotropic (c.f.~Fig.~\ref{fig1}), up to  the point where the applied
stress can no longer be  sustained, and the packing yields. While this
qualitative  picture has  been  long  established, due  to  a lack  of
appropriate  analytic  and numerical  tools,  quantitative studies  of
packings     under    static    shear     have    been    few
\cite{radjai98,geng03,majmudar05,atman05-2,snoeijer05}.

Recently,  a  novel  characterization  has  been  introduced  for  the
geometrical   patterns  formed  by   large  forces   in  isotropically
compressed force networks \cite {ostojic06}).  The patterns, displayed by molecular
dynamics (MD)  simulations of  granular packings, turned  out to have
scaling     properties     which     are    independent     of     the
pressure, polydispersity  and  friction.  An intriguing  question  is
whether an  analysis of scaling  properties can provide  insight into
the organization of force networks in packings under shear. This is a
much more difficult problem.

  Creating static  packings under shear  with MD is seriously
hindered by local rearrangements that  seem to prevent the system from
reaching  a clear  mechanical  equilibrium.  To  allow for  systematic
examinations  of   the  effects  of   shear,  it  has   been  proposed
\cite{snoeijer04-1,snoeijer05}  to  ignore the  microscopic  rearrangements of  the
grains, and study  the ensembles of force networks  allowed on a given
contact  network, as  function of  macroscopic stresses.   This purely
statistical approach was found to account well for the properties of
    packings under shear such as the existence of a yield threshold 
\cite{snoeijer05}, as  well as for the scaling  properties of clusters
of large forces in packings without shear \cite{ostojic06}. 
It also describes remarkably well the distribution of force magnitudes
\cite{snoeijer04-1,tighe05,snoeijer05} and the response to an external
overload \cite{ostojic06-2}.

In this Letter,  we study the geometry of patterns  of large forces in
two-dimensional, static packings under  shear generated from the force
network ensemble.  We find that applying shear stress leaves unchanged
the universal  scaling properties of the patterns,  but introduces two
different  length scales  in the  two directions  of  principal stress
axes.  We characterize  the anisotropy  of the  force networks  by the
ratio $r$ of these  length scales.  An anisotropic finite-size scaling
analysis allows us  to determine $r$ as function  of the applied shear
stress $\tau$ and the coordination number of the packing.  In contrast with
other properties such  as the yield stress, $r(\tau)$  turns out to be
independent of the underlying contact geometry, and thus provides a universal characterization of shear-induced anisotropy.

{\em Force networks---} To study systematically the influence of the
shear  stress  on the  geometry  of  large  forces we  examined  force
networks  generated  from the  force  network  ensemble pertaining  to
packings  of  frictionless,  non-attracting particles.  This  approach
relies on  the fact that in  a typical packing, the  number of unknown
contact forces is  larger than the number of  equations for mechanical
balance  \cite{snoeijer04-1}.  For a  given, fixed,  contact geometry,
the force network ensemble consists of all force networks with repulsive
contact forces in mechanical balance on each grain, and
consistent with the tensor $\sigma$ of applied macroscopic stress. The
considered  contact networks  are isotropic,  and the  coordinates are
such     that     the      pressure     is   isotropic,
i.e.~$\sigma_{xx}=\sigma_{yy}$,  while the dimensionless  shear stress
$\tau=\sigma_{xy}/\sigma_{xx}$ is varied.

 The ensemble is sampled numerically using two different methods : (i)
the wheel-move algorithm \cite{tighe05} for ordered, hexagonal packings
($z=6$); (ii) the  procedure outlined in \cite{snoeijer05} for
disordered packings of various  coordination numbers $z$. Both methods
  efficiently generate very large packings ($10^5$ and $5.10^4$
particles respectively), from which smaller sub-systems were extracted
and analyzed.   Note that  the ensemble is  non-empty only  for $\tau$
smaller than a maximal value $\tau_{max}(z)$, which has been identified as
the yield stress \cite{snoeijer05}.

{\em  Force chains---}  To study  the geometry  of patterns  formed by
large  forces, we  follow the  method introduced  in \cite{ostojic06},
which  we  here briefly  recall  and  then  extend to  packings  under
shear. The force network in a packing of frictionless particles can be
represented as a set of bonds connecting grains in contact, where each
bond  carries  the  magnitude   of  the  corresponding  contact  force
(c.f.~Fig.~\ref{fig1}  (a-c)).  A natural  way  to  isolate the  force
chains in  such a network is  to choose a threshold  $f$, and consider
only  the subgraph  formed by  bonds carrying  forces larger  than $f$
(c.f.~Fig.~\ref{fig1} (d-f)).  Varying this threshold  allows to study
patterns of large  forces at different scales: for  small values, most
of the grains remain connected, but as the threshold is increased, the
extracted subgraph  breaks into  disconnected clusters. The  extent of
each chain of forces larger than the threshold can be characterized by
the size of the corresponding cluster, quantified by the number $s$ of
mutually connected bonds.

In  an  ensemble  of   packings,  a  statistical  description  of  the
fluctuating patterns of large forces is given by the statistics of the
clusters obtained  at different  thresholds. Following the  methods of
percolation theory  \cite{stauffer:book}, the geometry  of an ensemble
of packings can be characterized by $P(s,f)$, the number (per bond) of
clusters  of  size  $s$  at  the threshold  $f$.  For  packings  under
isotropic pressure, it is moreover natural to describe the clusters by
a  single  characteristic   length,  the  cluster  correlation  length
$\xi(f)$, which corresponds to the  typical linear size of the 
clusters  at a  given threshold  (not considering  clusters percolating
through the whole system).

{\em Scaling analysis---}Analogously  to percolation and other lattice
models of critical phenomena, the  system is critical around the value
$f_c$ of  the threshold above which  no infinite cluster  is found. At
this  value,  $\xi(f)$  diverges  as  $|f-f_c|^{-\nu}$,  and  $P(s,f)$
becomes a  power-law, its moments  $\langle s^n \rangle$  diverging as
powers  of  $\xi(f)$:  $\langle  s^n \rangle\sim  \xi^{\phi_n}$.   For
systems of finite size, the correlation length is naturally bounded by
the  linear size  of  system, $L$, and  for  $\xi$ comparable  to $L$,  the
behavior of  $\langle s^n \rangle$ is  given by a  scaling function of
$L/\xi$ \cite{privman:book,binder89}:
\begin{equation}
\langle s^n(f,L) \rangle \sim L^{\phi_n}\Sigma_n[(f-f_c)L^{1/\nu}], \label{iso-fss}
\end{equation}
where  $\phi_n$  and  $\nu$  are  universal  critical  exponents,  and
$\Sigma_n$  universal  scaling functions.   For  force networks  under
isotropic  pressure,  $\phi_2=1.89$,  $\nu=1.65$,  and  $\Sigma_2$  is
independent  of  pressure,  polydispersity,  friction  and  force  law
\cite{ostojic06}.  From now on, we  will only consider the case $n=2$,
and omit the indices $n$ in the right-hand-side. Note that the scaling
function  $\Sigma_2$ varies  smoothly  and displays  a single  maximum
close to zero.

Equation  (\ref{iso-fss})  assumes that  the  size  of the examined
domains is isotropic. In the following, we will need to consider
  rectangular domains of size
$L_1\times   L_2$, in which case  the   scaling  function   $\Sigma$  also  depends   on the  aspect ratio  $L_2/ L_1$.   Considering  only the
maximum of $\langle s^2(f,L_1,L_2) \rangle$ with respect to $f$, which
is  equivalent  to looking  only  at  the  behavior  at  the effective
critical point, Eq.~(\ref{iso-fss}) reduces to
\begin{equation}
\langle s^2(L_1,L_2)\rangle _{max}=L_2^{\phi}{\bar \Sigma}\left(\frac{L_1}{L_2}\right).\label{eq:ch4-fss2}
\end{equation}
The scaling function ${\bar \Sigma}(x)$ simply expresses the fact
 that if $L_1\ll L_2$, the effective correlation length is set by
 $L_1$, so that $\langle s^2\rangle _{max}\sim L_1^{\phi}$,
 and correspondingly if $L_2\ll L_1$, $\langle s^2\rangle _{max}\sim
 L_2^{\phi}$. We therefore have (with $y=L_1/L_2$)
\begin{equation}
   {\bar \Sigma}(y) \sim\left\{
    \begin{array}{ll}
     y^{\phi} & \textrm{for $y \ll 1$}\\
     1 & \textrm{for  $y \gg 1$}
    \end{array}\right.
  \end{equation}
For $L_1\sim L_2$, there is a crossover between the two asymptotic trends, and the precise
behavior is determined by the scaling function.

{\em  Shear-induced  anisotropy---}  The  analysis  presented  so  far
pertains to the  case of isotropic pressure. In  packings under shear,
the force networks become increasingly anisotropic as the shear stress
is  increased,  the  large   forces  aligning  preferentially  in  the
direction  of  the   maximal  stress  axis  (cf.~Fig.~\ref{fig1}).  In
consequence,  close to the percolation threshold the clusters 
can  no longer be  described by  a single  length scale.   Instead two
correlation  lengths  $\xi_M$   and  $\xi_m$  must  be  distinguished,
 in the direction of  maximal and minimal stress axes respectively.  In
principle, at  criticality these two length scales  could diverge with
two different  exponents $\nu_M$  and $\nu_m$.  A  finite-size scaling
analysis as in  \cite{ostojic06}, but now varying independently
the  system sizes  $L_M$  and  $L_m$ in  the  directions of  principal
stresses, shows that $\nu_M=\nu_m=\nu$, the exponents $\phi$ and $\nu$
being equal to those of isotropically compressed packings.

Although  they diverge  with  the  same exponent,  the  values of  the
correlation lengths $\xi_M$ and  $\xi_m$ clearly differ.  This implies
that, to leading order,
\begin{equation}
\xi_{M}=\xi_{M}^{(0)}|f-f_c|^{-\nu} \,\,\,\mathrm{ and}\,\,\,
\xi_{m}=\xi_{m}^{(0)}|f-f_c|^{-\nu}. \label{eq:ch4-xi0}
\end{equation}
where the length scales  $\xi_{M}^{(0)}$ and $\xi_{m}^{(0)}$ depend on
the shear stress $\tau$ but not  on the threshold $f$. In the language
of critical phenomena, this  situation is called {\em weakly anisotropic
scaling},  as opposed  to the  case where  the  two correlation-length
exponents differ \cite{binder89}.

  To quantify  the anisotropy of  force networks as function  of shear
stress,  from Eq.~(\ref{eq:ch4-xi0}),  it appears  natural to  use the
anisotropy ratio $r(\tau)$ defined as
\begin{equation}
r(\tau)=\frac{\xi_{m}^{(0)}}{\xi_{M}^{(0)}}.
\end{equation}
Clearly, for $\tau=0$, $r=1$, and as $\tau$ is increased, $r$
decreases below one.

The anisotropy ratio $r$ as function of $\tau$ can be determined using
an anisotropic  scaling analysis. The central observation  is that, in
the context  of conventional critical phenomena,  a weakly anisotropic
system can  be made isotropic simply  by rescaling the  lengths in the
two  directions  of principal  axes.   In  the  present setting,  this
property  suggests that  if the  actual  lengths $L_M$  and $L_m$  are
replaced  by  properly rescaled  effective  lengths $\tilde{L}_M$  and
$\tilde{L}_m$, the  scaling properties  of the sheared  force networks
should   be   identical   to   those   of   isotropically   compressed
networks.  This  property  allows  us to  determine  the ratio of length scales
$\xi_{M}^{(0)}$ and $\xi_{m}^{(0)}$.
 
More specifically, the scaling of  the maximum of $\langle s^2\rangle$ is
described  by Eq.~(\ref{eq:ch4-fss2})  with an additional dependence on
the shear  stress $\tau$  in the right  hand side.  Our  hypothesis is
that the  scaling function ${\bar \Sigma}$ does  not depend explicitly
on  $\tau$,  and that  the  anisotropy  can  be eliminated  simply  by
replacing  $L_M$  and  $L_m$  by rescaled  lengths  $\tilde{L}_M$  and
$\tilde{L}_m$ given by
\begin{equation}
\tilde{L}_M =b_M(\tau) L_M \nonumber\,\,\, \mathrm{ and} \,\,\,
\tilde{L}_m=b_m(\tau) L_m  \label{eq:ch4-bs}
\end{equation}
with $b_M=1/\xi_{M}^{(0)}$  and  $b_m=
1/\xi_{m}^{(0)}$,                 and                therefore
$r(\tau)=b_M/b_m$.  Substituting  into Eq.~(\ref{eq:ch4-fss2}),  the
scaling relation becomes
\begin{equation}
\langle s^2(L_M,L_m,\tau)\rangle _{max}=(b_m L_m)^{\phi}{\bar
  \Sigma}\left(\frac{b_M L_M}{b_m L_m}\right).\label{eq:ch4-fss3}
\end{equation}
where the dependence on $\tau$ occurs only through the scale factors
$b_m$ and $b_M$.

\begin{figure}
\begin{center} 
\includegraphics [width=0.8\linewidth]{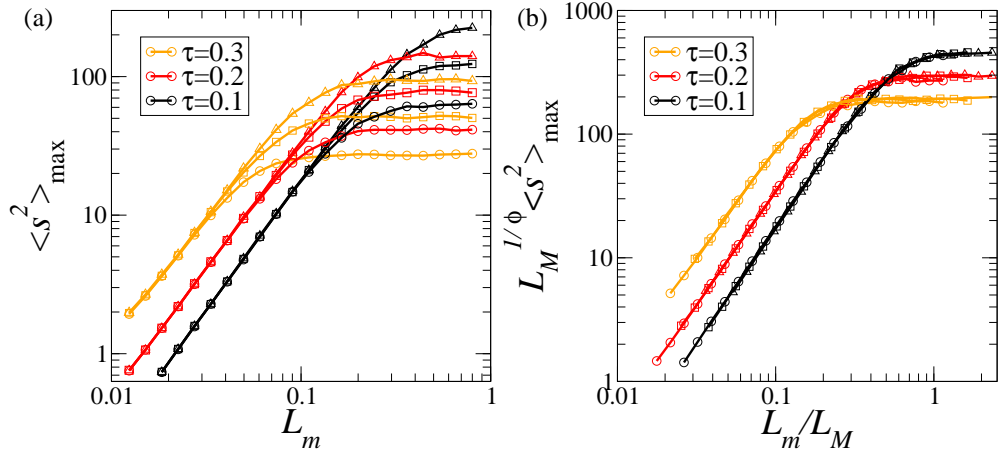}
\caption{(color online) Scaling of the maximum of the second moment of cluster sizes:
  (a)  $\langle s^2\rangle_{max}(L_M, L_m)$   as function  of $L_m$.   Three different  values of  $L_M$ are
  represented  with  circles,   squares  and  triangles,  while  three
  different values  of the shear  stress are shown in  three different
  colors.   The displayed  data  corresponds to  hexagonal packing  of
  disks,  i.e.~ $z=6$.  (b) Collapses  of data  corresponding  to same
  $\tau$,      obtained     by      rescaling     the      axes     by
  $L_M$. The curves are colored in the order of the legend. \label{fig:ch4-b4_collapse}} 
\end{center}
\end{figure}

The  validity  of  the  scaling relation  (\ref{eq:ch4-fss3})  can  be
checked          directly          from         the          numerical
data.  Fig.~\ref{fig:ch4-b4_collapse}   illustrates  the  behavior  of
$\langle s^2 \rangle_{max}$ as  function of $L_m$, for three different
values  of  $\tau$,  and  in  each  case  three  different  values  of
$L_M$.  For fixed $L_M$  and $\tau$,  as $L_m$  increases, at  first a
clear  power-law  can be  observed.  For  larger  values of  $L_m$,  a
crossover occurs and $\langle s^2 \rangle_{max}$ reaches a plateau. As
$L_M$ is increased,  the behavior in the scaling  regime is unchanged,
but the crossover occurs at larger $L_m$, and the value of the plateau
increases.   According to  Eq.~(\ref{eq:ch4-fss2}), the  value  of the
crossover  should scale  as $L_M$,  and the  value of  the  plateau as
$L_M^{\phi}$.   Rescaling  both   axes  appropriately,  the  data  for
different  $L_M$  indeed collapse  on  the  same  curves as  shown  in
Fig.~\ref{fig:ch4-b4_collapse}~(b).

For different values of the shear stress, the exponents in the scaling
regimes appear  to be identical,  however the prefactors of  the power
laws  clearly depend  on $\tau$.  From  Eq.~(\ref{eq:ch4-fss3}), these
prefactors correspond to $b_M$ and $b_m$, which can thus be determined
respectively from the  value reached at the plateau  and the intercept
of the  power law. Replacing  $L_M$ and $L_m$ by  rescaled lengths
$\tilde{L}_M$ and  $\tilde{L}_m$ defined in  Eq.~(\ref{eq:ch4-bs}), if
the scaling function ${\bar \Sigma}$ is independent of $\tau$, all the
data  must  collapse  on  a  single  curve. Note  that  as  $\tau$  is
increased, the cross  over occurs at smaller $L_m$,  and the domain of
power-law  scaling  shrinks,  so  that  $b_m$ can  be  extracted  only
for $\tau$ sufficiently smaller than the yield stress $\tau_{max}$ (beyond which the force network ensemble is empty).

\begin{figure*}
\begin{center} 
\includegraphics [width=0.45\linewidth]{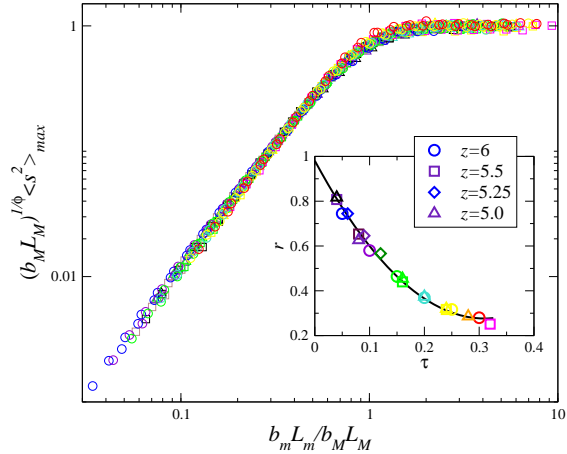}
\end{center}
\caption{(color online) The    anisotropic   scaling    function    ${\bar   \Sigma}$
  (cf.~Eq.~(\ref{eq:ch4-fss2})):   collapse  of   data   obtained  for
  different  values  of  shear  stress $\tau$  and  different  packing
  geometries. Each symbol   corresponds to data for a different
  coordination number  $z$ ($z=6$: regular,  hexagonal packing, already displayed in Fig.~
  \ref{fig:ch4-b4_collapse}; all others: disordered packings).  Each value of $\tau$ is represented
  by  a different  color.  The inset  shows  the anisotropy  parameter
  $r=\frac{b_M}{b_m}$ as  function of  $\tau$, for different $z$. The
  black solid line is a quadratic fit.\label{fig:ch4-tau_collapse}}
\end{figure*}

Fig.~\ref{fig:ch4-tau_collapse} displays the  results obtained for all
considered  values   of  shear  stress  $\tau$,   on  several  packing
geometries of different coordination  number $z$ ($z=6$ corresponds to
the   ordered   hexagonal  case,   while   the   other  packings   are
disordered). All  the data  clearly collapses on  the same  curve, in
agreement with the hypothesis 
that the scaling function  does not depend on $\tau$. As
postulated,  a  simple rescaling  of  the  length  scales in  the  two
directions  of  anisotropy is  thus  sufficient  to recover  isotropic
scaling in force networks under shear.The inset  of Fig.~\ref{fig:ch4-tau_collapse} displays  the anisotropy
parameter  $r=\frac{b_M}{b_m}$ as  function  of $\tau$.  Unexpectedly,
$r(\tau)$  appears to be  independent of  the coordination  number and
geometry of the
underlying  contact  network.

{\em Discussion---}  Our results show that applying  an external shear
stress  on a  force  network  does not  affect  the universal  scaling
properties of the force chains, but only induces two different length scales
in  the directions  of the  two  principal stress  axes.  While  these
typical lengths  and their ratio $r$  are not a priori  expected to be
universal,  we  find  that  they  are identical  for  various  contact
networks we  considered, which include the  regular hexagonal packing,
and disordered networks of different coordination numbers, and in this
sense universal.


An important remaining question is  the behavior close to the yielding
point. If yielding is analogous to a phase transition, as suggested by
the  jamming picture  \cite{liu98,jamming:book},  it could  be
expected that close to it a cross-over occurs, and the scaling
properties of clusters of  large forces change significantly. The
value of the yield stress $\tau_{max}$ was found to be strongly dependent on the
coordination number $z$ of the contact network \cite{snoeijer05}. On the other hand, we
find that $r(\tau)$ is completely independent of $z$ up to $\tau_{max}(z)$.
This observation suggests  the absence  of any diverging  or vanishing
length   scale  which   would  accompany   the  cross-over   close  to
$\tau_{max}$. Moreover,  we have not  observed any dramatic  change in
the scaling  properties close to $\tau_{max}$, but  additional work is
necessary to clarify these issues.

TJHV acknowledges financial support from the Netherlands Organization
for Scientific Research (NWO-CW) through a VIDI grant. SO is financially supported by the Dutch research organization FOM.

\end{document}